\documentstyle[aps]{revtex}
\begin{document}
\title{Cosmic Censorship: the Role of Quantum Physics}
\author{Shahar Hod}
\address{The Racah Institute for Physics, The
Hebrew University, Jerusalem 91904, Israel}
\date{\today}
\maketitle

\begin{abstract}
The cosmic censorship hypothesis introduced by Penrose thirty years
ago is still one of the most important open questions in {\it classical} general
relativity. The main goal of this paper is to put forward the idea
that cosmic censorship is intrinsically a {\it quantum} phenomena. We
construct a gedanken experiment which seems to violate the cosmic
censorship principle within the purely {\it classical} framework of general
relativity. We prove, however, that {\it quantum} physics restores the
validity of the conjecture. It is therefore suggested that cosmic
censorship might be enforced by a quantum theory of gravity.
\end{abstract}
\bigskip

\section{Motivation and Objectives}\label{Sec1}

It is well-known that atoms are unstable
objects if viewed in the framework of the purely {\it classical} laws.
The stability of atomic systems, and therefore our very existence 
depends in an essential manner on a {\it quantum} effect, namely the 
Heisenberg quantum uncertainty principle.

In some respects the black hole 
plays the same role in gravitation that the atom played in the early development of quantum
mechanics \cite{Beken1}. The
quantum nature of black holes manifests itself most vividly in the
form of a discrete eigenvalue spectrum for the 
black-hole {\it horizon} area \cite{Hod1}. 
Thus, it is not at all surprising that the stability 
of the black-hole horizon might also be 
a {\it quantum} phenomena, basically not different from the stability of
the atom. Attempts to establish a {\it general} proof for the stability of the
black-hole event horizon in the framework of the purely classical laws may therefore
seem quite presumptuous. 

The main goal of this paper is to put forward the idea that the
stability of the black-hole horizon, and therefore 
the {\it cosmic censorship} principle are intrinsically {\it quantum}
phenomenon \cite{note1}. To that
end, we shall construct a gedanken experiment which seems to violate the cosmic
censorship hypothesis within the purely {\it classical} framework of general
relativity. We will prove, however, that in accordance with our
conjecture {\it quantum} effects
restore the cosmic censorship principle to its proper status.

\section{A Brief Review of Former Gedanken Experiments}\label{Sec2}

Spacetime singularities arising from realistic gravitational collapse scenarios are
always hidden inside of black holes, and are therefore invisible to
distant observers. This is the essence of the (weak) cosmic censorship hypothesis, 
put forward by Penrose \cite{Pen} thirty years ago. The conjecture is
widely believed to be true and has become one of
the corner stones of classical general relativity. 
However, despite the flurry of activity over the years, we are 
still lack a general proof of this conjecture. 

When a rigorous mathematical proof seems to be beyond our present reach, it
may seems quite tempting to obtain a convincing counterexample and show
that the conjecture is false. If the conjecture fails, 
then it is quite possible that the formation of
a black hole would be a non-generic outcome of gravitational
collapse. {\it If} so, one might expect to find some evidence for an 
instability of the black-hole event horizon in physical
processes which seem to have a chance of exposing the singularity hidden inside a
black hole. For the advocates of the cosmic
censorship principle the task remains to find out how such candidate
processes eventually fail to remove the horizon.

We shall not attempt to review the numerous works that have been
written addressing the question of whether or not the cosmic
censorship hypothesis holds (for some of the recent
reviews and lists of references, see e.g. \cite{Wald1,Sin}). 
Rather, we will briefly describe those 
works which are directly related to the present paper. 

One of the earliest attempts to eliminate the horizon of a black hole
is due to Wald \cite{Wald2}. As is well-known, the Reissner-Nordstr\"om metric 
with $M < Q$ (where $M$ and $Q$ are the mass and charge of the
configuration) does not contain an event
horizon, and therefore describes a naked singularity. One may start
with an extremal black hole (characterized by $Q=M$), and try to
``supersaturate'' the extremality condition by dropping in a test charged
particle whose charge-to-mass ratio is greater than unity. Wald
\cite{Wald2} showed, however, that such an attempt would fail because of the
Coulomb potential barrier surrounding the black hole. 

Hiscock \cite{His}, and independently Bekenstein and Rosenzweig
\cite{BekRos} attempted to overcharge a black hole in a different 
version of the gedanken experiment: suppose there exist two different types
of local charge, namely type-$q \in U(1)$ and type $k \in U'(1)$, e.g.,
electric and magnetic charge. The black hole is assumed to be an 
extremal Reissner-Nordstr\"om black hole, possessing a $U'(1)$
charge, but no $U(1)$ charge. Thus, the black hole is not
endowed with a $U(1)$ gauge field, and an infalling 
charge $q \in U(1)$ seems to encounter no repulsive electrostatic potential
barrier. 

Bekenstein and Rosenzweig \cite{BekRos} considered the specific case
of a charged particle which starts falling from spatial {\it infinity} (thus, the particle's
energy-at-infinity is larger than its rest mass). It was shown in
\cite{BekRos} that such an attempt to
overcharge a black hole would fail because the required classical radius of the charged
body (the analogous of the well-known classical radius of the
electron) is larger than the black-hole size.

The natural question immediately arises: what physical mechanism insures the
stability of the horizon if the charged particle is {\it slowly} lowered 
towards the black hole ? In this case, the energy delivered
to the black hole can be {\it red-shifted} by letting the
assimilation point approach the black-hole horizon. At first sight,
therefore, the particle is not hindered from entering the black hole and removing
its horizon, thus violating cosmic censorship. The final outcome of
this `dangerous' gedanken experiment has remained unclear for
almost two decades. Recently, Hod \cite{Hod2} has reexamined this old
question and showed that this process actually fails to remove the
horizon; the black hole preserves its integrity thanks to two factors 
not considered in former gedanken 
experiments: the effect of the spacetime curvature on the 
electrostatic {\it self-interaction} of the charged body (the 
black-hole polarization), and the {\it finite} size imposed on a charged
body which respects the weak (positive) energy condition. 

Quinn and Wald \cite{QuWa} have independently suggested to use the
self-energy correction in the context of the gedanken experiment
recently proposed by Hubeny \cite{Hub}. Perhaps somewhat surprisingly,
it was shown in \cite{Hub} that the test particle
approximation actually allows a {\it near} extremal black hole to ``jump over'' extremality by
capturing a charged particle which starts falling from spatial infinity. The
({\it classical}) effect of the self-energy turns over this conclusion.

In this paper we inquire into the physical mechanism which protects
the black-hole horizon from being eliminated by the assimilation of a
charged object which is {\it slowly} lowered into a (near extremal)
black hole (this is the more `dangerous' version of the original
gedanken experiment \cite{Wald2,Hub}). We will prove that 
purely {\it classical} effects are actually helpless against the
exposure of a naked singularity in this gedanken experiment. 
However, we shall propose a resolution out of this `embarrassing' situation which involves 
the {\it quantum} properties of the vacuum.

\section{The Gedanken Experiment}\label{Sec3}

We consider a charged body of rest mass $\mu$, charge $q$, and proper
radius $b$, which is {\it slowly} descent into a (near extremal) black hole. 
The total energy $\cal E$ of the body in a black-hole spacetime 
is made up of three contributions: $ 1) \  {\cal E}_0$, the energy
associated with the body's mass 
(red-shifted by the gravitational field); $ 2)  \ {\cal E}_{elec}$, the
electrostatic interaction of the charged body with the external electric
field; and $ 3) \ {\cal E}_{self}$, the gravitationally induced self-energy of the charged body. 

The first two contribution, ${\cal E}_0+{\cal E}_{elec}$, are given by Carter's \cite{Carter}
integrals of the Lorentz equations of motion for a charged particle
moving in a charged black-hole background \cite{note2}:

\begin{equation}\label{Eq1}
{\cal E}_0+{\cal E}_{elec}={{\mu \ell (r_+-r_-)} \over {2{r^2_+}}} 
[1+O(\ell^2/{r^2_+})] +{{qQ} \over {r_{+}}}
-{{qQ\ell^2(r_{+}-r_{-})} \over {4{r^4_+}}}[1+
O(\ell^2/{r^2_+})]\  ,
\end{equation}
where $r_{\pm}=M \pm (M^2-Q^2)^{1/2}$ are the
locations of the black-hole (event and inner) horizons 
(we use gravitational units in which $G=c=1$), and 
$\ell$ is the proper distance from the horizon. Namely,

\begin{equation}\label{Eq2}
\ell = \ell(r)= \int_{r_{+}}^{r} \sqrt{g_{rr}} dr\  ,
\end{equation}
with $g_{rr}=r^2/(r-r_+)(r-r_-)$. 

The third contribution, ${\cal E}_{self}$, reflects the effect of the 
spacetime {\it curvature} on the particle's 
electrostatic {\it self-interaction}. The physical origin of this force 
is the distortion of the charge's long-range Coulomb field by
the spacetime curvature. This can also be interpreted as being due to the
image charge induced inside the (polarized) black hole
\cite{Linet,BekMay}. The self-interaction of a charged particle in the
black-hole background results with a repulsive (i.e., directed away from the
black hole) self-force. A variety of
techniques have been used to demonstrate this effect in black-hole 
spacetimes \cite{DeDe,Ber,Mac,Vil,SmWi,ZelFro,Lohi,LeLi1,LeLi2}. In particular, 
the contribution of this effect to the
particle's (self) energy in the Reissner-Nordstr\"om 
background is ${\cal E}_{self}=Mq^2/2r^2$ \cite{ZelFro,Lohi}, which implies
${\cal E}_{self}=Mq^2/2{r^2_+}$ to leading order in $(\ell /r_+)^2$. 

We thus obtain

\begin{equation}\label{Eq3}
{\cal E}(\ell)={{\mu \ell (r_+-r_-)} \over {2{r^2_+}}} +{{qQ} \over {r_{+}}}
-{{qQ\ell^2(r_{+}-r_{-})} \over {4{r^4_+}}}+ {{Mq^2} \over {2{r^2_+}}}\  .
\end{equation}
This expression is actually the effective potential governing the
motion of a charged body in the black-hole background. Provided
$qQ>0$, it has a {\it maximum} located at $\ell=\ell^*(\mu,q;M,Q)=\mu r^2_+ /qQ$. 
The charged body has to be over this 
potential barrier in order to be captured by the black hole. 

The gradual approach to the black hole must stop when the
proper distance from the body's center of mass to the black-hole
horizon equals $b$, the body's radius. For bodies which satisfy 
the restriction $b \leq \ell^*$ one should therefore 
evaluate $\cal E$ at the point $\ell=b$. 
An assimilation of the charged object results with a 
change $\Delta M={\cal E}$ in the
black-hole mass and a change $\Delta Q=q$ in its charge. The condition
for the black hole to preserve its integrity after the assimilation of
the body is therefore

\begin{equation}\label{Eq4}
q+Q \leq M+{\cal E}\  .
\end{equation}
Substituting ${\cal E}={\cal E}_0+{\cal E}_{elec}+{\cal E}_{self}$
from Eq. (\ref{Eq3}) we find a necessary and
sufficient condition for
removal of the black-hole horizon:

\begin{equation}\label{Eq5}
(q-\varepsilon)^2+{{2\varepsilon} \over M} \Bigg (\mu b -q^2 -{{qb^2} \over
  {2M}} \Bigg) +{{q\varepsilon^2} \over M} < 0\  ,
\end{equation}
where $r_{\pm} \equiv M \pm \varepsilon$. This condition is
accurate to order $O(\varepsilon^2)$. The expression on the l.h.s. of
Eq. (\ref{Eq5}) is minimized
for $q = \varepsilon +O(\varepsilon^2/M)$, thus yielding

\begin{equation}\label{Eq6}
2\mu b -q^2 -qb^2/M <0\  ,
\end{equation}
as a necessary and sufficient condition for elimination of the black-hole horizon.

The total mass of the charged body is given by $\mu =\mu_0+ fq^2/b$,
where $\mu_0$ is the mechanical (nonelectromagnetic) mass, and $f$ is
a numerical factor of order unity which depends on how the charge is
distributed inside the body. The Coulomb energy 
attains its {\it minimum}, $q^2/2b$, when the charge is uniformly spread on a thin
shell of radius $b$, which implies $f \geq 1/2$ (an homogeneous
charged sphere, for instance, has $f=3/5$). Therefore, any charged body which respects the
weak (positive) energy condition must be larger than $r_c \equiv
q^2/2\mu$. 

In deriving the lower bound on particle's size, $r_c$, 
one neglects the mechanical mass of
the body. In fact, large stresses may be placed inside the charged
body and the charge distribution must have forces of nonelectromagnetic
character holding it stable. Therefore, a purely classical
electromagnetic model has little relevance to the {\it real} world. Nonelectromagnetic forces imply a 
large contribution $\mu_0$ to the mass of the body from such forces. The
large nonelectromagnetic contribution will
prevent us from getting close to the minimal size limit $r_c$; Atomic
nuclei, for instance, are bounded by {\it strong} forces,
which are often much stronger than the force exerted by the surface
electric field. In fact, even atomic nuclei, which are 
the {\it densest} charged objects (with negligible self-gravity) in
nature, satisfy the relation $b/{r_c} \sim
10^2-10^3$ and are therefore {\it far larger} than $r_c$ ! Black holes
with their extreme {\it gravitational} binding character are in fact the only
objects in nature whose size can come close to the limit $r_c$: An 
extremal Reissner-Nordstr\"om black hole, in particular, 
satisfies the relation $b/{r_c}=2$ (other black holes satisfy
$b/{r_c}>2$). Therefore, even the gravitational interaction in its
extremal form as displayed in black holes {\it cannot} allow a charged
object to be as small as $r_c$. 

Thus, one may safely conjecture that a charged body which respects the
weak (positive) energy condition
must satisfy the restriction $b/{r_c} \geq 2$ (where the equality
is only saturated by the extremal Reissner-Nordstr\"om black
hole). This result, combined with the inequality $b \le \ell^*$
(i.e., $\mu b \ge qb^2/M$) implies that 
$2\mu b -q^2 -qb^2/M \geq 0$. We therefore conclude that the
black-hole horizon cannot be
removed by an assimilation of such a charged body -- cosmic
censorship is upheld !

For an elementary charge which is subjected to Heisenberg's quantum
uncertainty principle with $b \sim \hbar/ \mu$, 
$r_c$ is not the measure of particle size. In fact, for $U(1)$ charges found free in nature
(weak coupling constant $q^2 \ll \hbar$, e.g., an electron), the classical radius $r_c$
is far smaller than the Compton length. This is incompatible with the necessary
condition Eq. (\ref{Eq6}), and we therefore recover our
previous conclusion that the black-hole horizon cannot be removed. 

Charged bodies which satisfy the relation $b > \ell^*$ must have a
minimal energy of ${\cal E}_{min}={\cal E}(\ell^*)$ in order to overcome the
potential barrier, and to be captured by the black hole (this is also
true for any charged object which is released to fall in from 
$\ell  > \ell^*$). Taking
cognizance of Eq. (\ref{Eq6}) we find that a necessary and sufficient condition for
removal of the black-hole horizon is 
$2\mu \ell^* -q^2-q\ell^{*2}/M<0$, or equivalently,

\begin{equation}\label{Eq7}
{\mu^2}/{q^3} < E\  ,
\end{equation}
where $E = Q/{r^2_+}=M^{-1}+O(\varepsilon /M^2)$ is the
black-hole electric field in the vicinity of its horizon.

Is there a physical mechanism which restrains the black-hole electric field 
from growing beyond the `dangerous' value given in Eq. (\ref{Eq7}) ? 
The answer seems to be negative within the framework of the purely
{\it classical} laws. Note that the two series of inequalities $q^2/\mu \leq b \ll
r_+ \leq E^{-1} < q^3/{{\mu}^2}$ and $\mu/qE \equiv \ell^* < b \ll
r_+ \leq E^{-1}$ are easily satisfied by
charged objects with $\mu \ll q$. 
Thus, without any obvious physical mechanism which bounds the
black-hole electric field, the assimilation of a charged object by a
charged black hole which satisfies the condition Eq. (\ref{Eq7}) is
expected to {\it violate} the cosmic censorship conjecture.

This physical picture changes, however, in the framework 
of the {\it quantum} laws because now there is a physical mechanism
which bounds the electric-field strength -- {\it Vacuum polarization} effects
(Schwinger discharge of the black hole) do set an 
upper bound to the black-hole electric field ! 

A variety of techniques have been used to calculate the rate of
particle production by a constant electric field in flat
\cite{SHS} and curved \cite{MZCGPDN} spacetimes. 
The critical electric field, $E_c(\mu,q)$, for pair-production of particles 
with rest mass $\mu$ and charge $q$ is found 
to be $E_c(\mu,q)={\pi \mu^2} /{q \hbar}$. 
This order of magnitude can easily be understood 
on physical grounds: In a quantum theory the vacuum is continuously
undergoing fluctuations, where a pair of ``virtual'' particles is
created and then annihilated. The electric field tends to separate the
charges. If the field is strong enough, the particles tunnel through
the quantum barrier and materialize as real particles. Schwinger-type 
pair-production is therefore exponentially suppressed unless the work done by 
the electric field on the virtual pair of (charged) particles 
in separating them by a Compton wavelength
is (at least) of the same order of magnitude of the particle's mass. 

In practice, the electric field of
a black hole is bounded {\it quantum} mechanically by pair-production 
of the {\it lightest} charged particles in nature. This implies 
$E \leq E_c \equiv {{\pi {m^2_e}}/{|e| \hbar}}$, 
where $m_e$ and $e$ are the rest mass and charge of the electron,
respectively. We thus conclude that a necessary and sufficient condition for 
a violation of the cosmic censorship conjecture within the framework of 
a {\it quantum} theory is the existence in
nature of a charged object which satisfies the inequality

\begin{equation}\label{Eq8}
q^3E_c/{\mu^2} > 1\  .
\end{equation}
Obviously, the most dangerous threat to the integrity of the black
hole is imposed by the electron, which has the largest charge-to-mass 
ratio in nature. However, even the electron itself satisfies
the relation $q^3E_c/{\mu^2}=\pi \alpha < 1$ 
(where $\alpha=e^2/\hbar \simeq 1/137$ is the fine structure
constant), and thus it {\it cannot} remove the black-hole horizon. 
Atomic nuclei, the densest composite charged objects in nature 
satisfy the relation $q^3E_c/{\mu^2} \lesssim 10^{-7}$ 
and are therefore absolutely harmless to the black hole. Thus, we 
conclude that the mechanism of vacuum polarization (Schwinger
discharge of the black hole) insures the integrity of the black
hole. This quantum phenomena therefore prevents an exposure of a naked singularity.

\section{Conclusions}\label{Sec4}

Motivated by the plausible analogy between black holes and atoms we
have conjectured that the stability of the black-hole event horizon,
like the stability of the atom, is intrinsically a {\it quantum}
phenomena. To prove this conjecture we have constructed a gedanken
experiment which may be regarded as the most `dangerous' one among a
large family of gedanken experiments \cite{Wald2,His,BekRos,Hod2,Hub}
that have been designed to challenge the validity of the cosmic
censorship hypothesis. It was shown that purely {\it classical} effects are 
actually helpless against the exposure of a naked singularity when a 
charged object is assimilated by a black hole. We have demonstrated 
explicitly, however, that quantum effects help save cosmic censorship.

Although the question of whether cosmic censorship holds
remains very far from being settled, the physical picture that 
now arises can be summarized by the following two statements:
\begin{itemize}
\item The black-hole event horizon may be {\it classically} unstable while
absorbing charged objects. This suggests that the purely classical laws
of general relativity do not enforce cosmic censorship.
\item The stability of the black-hole event horizon depends in an 
essential manner on {\it quantum} effects. This 
suggests that cosmic censorship might be enforced by a quantum theory
of gravity.
\end{itemize}

We find it most intriguing that {\it quantum} effects must be invoked
in order to insure the stability of the black-hole event horizon, 
and hence to restore the validity of the cosmic censorship
principle \cite{note3}. We thus conclude that the cosmic censor must be 
cognizant of quantum physics.
 
\bigskip
\noindent
{\bf ACKNOWLEDGMENTS}
\bigskip

It is a pleasure to thank Jacob D. Bekenstein, Tsvi Piran, and Avraham
Gal for stimulating discussions. I also wish to
thank Veronika E. Hubeny for a correspondence. 
This research was supported by a grant from the Israel Science Foundation.


\begin{thebibliography}{99}

\bibitem{Beken1} J. D. Bekenstein, 
in {\it Proceedings of the VIII Marcel Grossmann Meeting on General Relativity}, 
edited by T. Piran and R. Ruffini (World Scientific, Singapore, 1998).

\bibitem{Hod1} See e.g., S. Hod, Phys. Rev. Lett. {\bf 81}, 4293
  (1998), and references therein.

\bibitem{note1} The idea to use quantum effects in order to insure the
  stability of the black-hole event horizon was first suggested in the
  seminal work of Bekenstein and Rosenzweig \cite{BekRos}, in which an
  elementary $U(1)$
  charged particle with $q^2 > \hbar$ is thrown into a
  Reissner-Nordstr\"om black hole with a {\it different} type of
  $U(1)$ charge. There are, however, two major shortfalls in this
  gedanken experiment which make this example not very convincing
  (although it is surely indicative): The discussion given in \cite{BekRos} is only qualitative
  because the analysis of a strongly coupled $U(1)$ theory is not yet
  feasible. More important, an elementary charged particle 
  with $q^2 >  \hbar$ actually violates the positive energy condition, which is a
  classical condition. Therefore, the very existence of such 
  particles is a {\it quantum} phenomena, and thus it is {\it a priori}
  obvious that quantum physics {\it must} be used in order to protect
  the black-hole horizon against such particles. In this paper we
  shall prove, perhaps somewhat surprisingly, that quantum effects must be invoked in order to
  preserve the black-hole integrity while absorbing absolutely
  `innocent' objects, i.e., charged bodies which do respect the {\it classical} energy condition.

\bibitem{Pen} R. Penrose, Riv. Nuovo Cimento {\bf 1}, 252 (1969);
  R. Penrose in 
{\it General Relativity, an Einstein Centenary Survey},
eds. S. W. Hawking and W. Israel (Cambridge University Press, 1979). 

\bibitem{Wald1} R. M. Wald, ``Gravitational Collapse and Cosmic
  Censorship'', e-print gr-qc/9710068.

\bibitem{Sin} T. P. Singh, ``Gravitational Collapse, Black Holes and
  Naked Singularities'', e-print gr-qc/9805066.

\bibitem{Wald2} R. Wald, Ann. Phys. (N. Y.) {\bf 83}, 548 (1974).

\bibitem{His} W. A. Hiscock, Ann. Phys. (N. Y.) {\bf 131}, 245 (1981).

\bibitem{BekRos} J. D. Bekenstein and C. Rosenzweig, Phys. Rev. D {\bf 50}, 7239 (1994).

\bibitem{Hod2} S. Hod, Phys. Rev. D (to be published).

\bibitem{QuWa} T. C. Quinn and R. M. Wald, e-print gr-qc/9903014.

\bibitem{Hub} V. E. Hubeny, Phys. Rev. D {\bf 59}, 064013 (1999).

\bibitem{Carter} B. Carter, Phys. Rev. {\bf 174}, 1559 (1968).

\bibitem{note2} Our task is to challenge the validity of the 
cosmic censorship conjecture in 
the most `dangerous' situation, i.e., when the charge-to-energy ratio of
the particle is as large as possible. Therefore, we consider a body which is 
captured from a radial turning point of its motion. 
This {\it minimize} the energy delivered to the
black hole (for a given charge of the body).

\bibitem{Linet} B. Linet, J. Phys. A: Math. Gen. {\bf 9}, 1081 (1976).

\bibitem{BekMay} J. D. Bekenstein and A. E. Mayo, e-print gr-qc/9903002.

\bibitem{DeDe} C. M. DeWitt and B. S. DeWitt, Physics (N. Y.) {\bf 1}, 3 (1964).

\bibitem{Ber} F. A. Berends and R. Gastmans, Ann. Phys. (N. Y.) {\bf 98}, 225 (1976).

\bibitem{Mac} C. H. MacGruder III, Nature (London) {\bf 272}, 806 (1978).

\bibitem{Vil} A. Vilenkin, Phys. Rev. D {\bf 20}, 373 (1979).

\bibitem{SmWi} A. G. Smith and C. M. Will, Phys. Rev. D {\bf 22}, 1276
  (1980).

\bibitem{ZelFro} A. I. Zel'nikov and V. P. Frolov, Sov. Phys. -JETP
  {\bf 55}, 191 (1982).

\bibitem{Lohi} D. Lohiya, J. Phys. A: Math. Gen. {\bf 15}, 1815 (1982).

\bibitem{LeLi1} B. L\'eaut\'e and B. Linet, J. Phys. A: Math. Gen. {\bf 15}, 1821 (1982).

\bibitem{LeLi2} B. L\'eaut\'e and B. Linet, Int. J. Theor. Phys. {\bf 22}, 67 (1983).

\bibitem{SHS} F. Sauter, Z. Phys. {\bf 69}, 742 (1931); 
W. Heisenberg and H. Euler, Z. Phys. {\bf 98}, 714 (1936); J. Schwinger, Phys. Rev. {\bf 82}, 664 (1951).

\bibitem{MZCGPDN} M. A. Markov and V. P. Frolov, Teor. Mat. 
Fiz. {\bf 3}, 3 (1970); W. T. Zaumen, Nature {\bf 247}, 531 (1974); 
B. Carter, Phys. Rev. Lett. {\bf 33}, 558 (1974); 
G. W. Gibbons, Comm. Math. Phys. {\bf 44}, 245 (1975); 
L. Parker and J. Tiomno, Astrophys. Journ. {\bf 178}, 809 (1972); 
T. Damour and R. Ruffini, Phys. Rev. Lett. {\bf 35}, 463 (1975); 
I. D. Novikov and A. A. Starobinsky, 
Zh. Eksp. Teor. Fiz. {\bf 78}, 3 (1980) [Sov. Phys. JETP {\bf 51}, 3 (1980)].

\bibitem{note3} An analogous conclusion had been considered to hold true
  in the context of the instability
  of Cauchy horizons inside charged black holes embedded in de
  Sitter spacetimes. It was believed that these inner horizons may be
  classically {\it stable}, thus violating strong cosmic
  censorship (see \cite{Poi} and references therein for additional details). 
  Quantum effects have been suggested in order to insure
  the {\it instability} of these Cauchy horizons, and thus to enforce
  cosmic censorship \cite{Poi}. However, Brady et. al. \cite{BME} have
  recently shown that these Cauchy
  horizons are actually always {\it classically} unstable. This restores the
  full validity of strong cosmic censorship within the purely classical
  framework of general relativity, and makes it unnecessary to invoke quantum effects.

\bibitem{Poi} E. Poisson, in {\it Internal structure of black holes and
  spacetime singularities, Volume XIII of the Israel Physical Society},
  Edited by L. M. Burko and A. Ori (Institute of Physics, Bristol, 1997).

\bibitem{BME} P. R. Brady, I. G. Moss and R. C. Myers,
  Phys. Rev. Lett. {\bf 80}, 3432 (1998).

\end{thebibliography}
\end{document}